# Experimental study of a tunable hybrid III-V-on-silicon laser for spectral characterization of Fiber Bragg Grating sensors.

Jean-Baptiste Quelene*[a], Didier Pohl[b], David Bitauld[c], Karim Hassan[d], Guillaume Laffont [a]

[a]Univ. Paris-Saclay, CEA, List, 91120 Palaiseau, France, [b]Safran S3AR, 1 r Jeunes Bois, 78117 Chateaufort, [c] III-V Lab, joint lab of Nokia, Thales & CEA, 1 A. Fresnel, 91767 Palaiseau, [d] CEA LETI, Minatec, 17 rue des Martyrs, F-38054 Grenoble

## ABSTRACT

Fiber Bragg Grating (FBG) sensors offer multiple benefits in comparison with electronic sensors due to their compactness, electromagnetic immunity as well as their resistance to harsh environments and their multiplexing capabilities. Structural Health Monitoring (SHM) is one of the various potential industrial applications that could take full advantage of those sensors. However, there is a need for a low size, weight, power and cost interrogation unit for certain application areas such as aerospace or aeronautics. That is the reason why recent efforts have been made to use integrated components and circuits for interrogation of FBGs. Among different techniques, interrogation with a swept laser source is of high interest since it has a high multiplexing capability and could reach a high level of integration using other integrated components such as photodetectors, grating couplers or directional couplers to form a compact interrogation unit. In this paper, we present characterization results of a fully-packaged hybrid III-V on silicon tunable laser diode operating in the C and L bands. Wavelength maps are produced and analyzed and modulation of emitted wavelength is discussed. Preliminary results corresponding to a moderate frequency (10-Hz sweep rate) were obtained and FBG reflection spectra acquired with a broadband source (BBS) and a swept laser diode are compared. Finally, we discuss potential design improvements in order to reach high scan rates (> 10 kHz) and a large tuning range.

## 1. INTRODUCTION

Fiber optic sensing has been investigated for decades for applications in various industrial fields ranging from structural health monitoring (SHM), oil & gas, as well as biomedical applications. Indeed, fiber-optic sensors such as Fiber Bragg Gratings (FBGs) are of particular interest since they benefit from their low electromagnetic immunity, high level of integration, high wavelength division multiplexing capabilities and resistance to harsh environments such as high temperatures (>600 °C) and irradiation. Research is being conducted in the aeronautic industry in order to provide fiber optic sensors and interrogation systems compatible with its requirements. Specifically developed coatings and fibers aiming at overcoming temperature limitations could potentially allow for temperature, strain or pressure measurements deep inside aircraft engines. However, challenges regarding the size, weight, power and cost of the interrogation unit remain in order to fulfill requirements for this field of applications and efforts have been made in order to integrate such optoelectronic systems using Photonic Integrated Circuits (PICs). Commercial units using a broadband source and PICs include specifically designed Array Waveguide Gratings (AWGs) on InP platform[1] or interrogators based on Planar Lightwave Circuits (PLC) with dispersive filter or AWG[2]. However, such interrogation techniques are limited in terms of Wavelength Division Multiplexing (WDM) capability and investigations have been undertaken for other interrogation schemes, such as interrogation with a swept laser source[3] that could allow for a higher number of sensors per channel and interrogation of other fiber optic sensors like Fabry-Perot cavities.

The advent of WDM in optical communications increased the interest for widely tunable laser sources able to address a number of WDM channels by tuning its wavelength over a large wavelength band. In particular, hybrid III-V- on-Si have experienced notable progress over the last decade[4], taking advantage of III-V material for optical gain and SOI platform for laser cavity mirrors and tuning mechanism. A tuning range as high as 90 nm (C+L bands) was demonstrated[5] with such a hybrid source including two micro-ring resonators in Vernier configuration and a phase section thermally tuned with NiCr heaters. Indeed, adjusting the power dissipated in one ring resonator will shift its transmission spectrum and consequently select different wavelength bands in accordance with the Vernier effect.

*jean-baptiste.quelene@cea.fr; cea.fr

However, continuous tuning of the emitted wavelength is required for FBG interrogation rather than a selectable one, and an experimental study of such a hybrid source with a lower tuning range is carried out in this paper in order to evaluate feasibility of wavelength sweeping over the full tuning range of the device. First, the tuning mechanism of the source is investigated through the acquisition and analysis of wavelength maps (2). Then, modulation of emitted wavelength by tuning both Vernier filters and the phase section is exposed in 3). Finally, potential design improvements are discussed in 4) in order to reach high sweep rates (> 10 kHz) and a larger continuous tuning range.

## 2. ACQUISITION AND REPRESENTATION OF WAVELENGTH MAPS

The first step of this investigation consists in the acquisition and representation of emitted wavelength maps as a function of electrical power dissipated in both Vernier filters. The setup includes a homemade electronic board delivering controlled output currents injected into heaters, a laser diode controller (ILX Lightwave LDC3722) that provides bias current for laser gain and temperature control of the chip as well as an Optical Spectrum Analyzer (OSA, Anritsu MS9740A) that enables analysis of the emitted spectrum. The power dissipated in both heaters is varied over a range of more than 70 mW by steps of a few hundreds of µW. At each step, the dominant emitted wavelength as well as the Side Mode Suppression Ratio (SMSR) are recorded. The following paragraph describes and analyzes such maps.

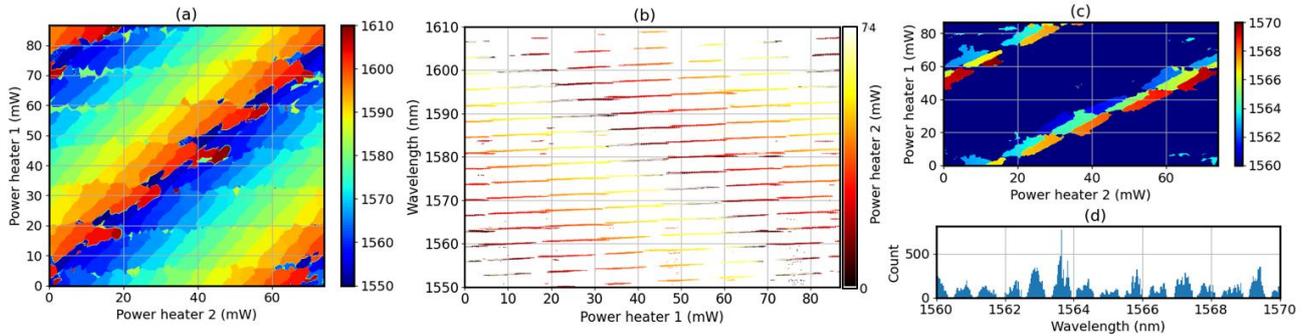

Figure 1. (a) Wavelength map of laser. (b) Projection of the map on the ($P_1$, $\lambda$) plane. (c) Wavelength map with limited wavelength range (1560-1570 nm). Values out of this range are set to zero. (d) Histogram of map (c).

Figure 1(a) depicts a wavelength map acquired with the laser diode for an OSA operating in the range of 1550–1610 nm with 30 pm resolution. One can observe that by raising the dissipated power $P_1$ in ring 1 heater (resp., $P_2$ in ring 2 heater), the emitted wavelength is decreased (resp., increased) by steps of ~3 nm until initial wavelength band is recovered when reaching a 2π-phase shift corresponding to ~70 mW. The total tuning range of the Vernier filters is $\Delta\lambda_{Vernier}$ ~50 nm. When projecting values on the plane ($P_1$, $\lambda$), as depicted on Figure 1(b), one can observe that data points are arranged in segments which slopes correspond approximately to ring heater 1 efficiency (~52 pm/mW) and which offset difference correspond to ring 1 Free Spectral Range (FSR) ~3.6 nm. The same operation can be done on the ($P_2$, $\lambda$) plane which enabled to retrieve an approximation of FSR of ring 2 (~3.4 nm) and its heater efficiency (~ 54 pm/mW). The value of $\Delta\lambda_{Vernier}$ calculated from those values and the well-known formula is in accordance with the value obtained directly from the map.

Figure 1(c) represents the same wavelength map with reduced scale (1560-1570 nm) and with points with wavelength out of this range set to zero. This representation allows for visual evaluation of potential paths on this map for a 10-nm sweep that consist of different diagonal segments. However, one can observe that continuous tuning of the Vernier through an entire segment is not possible since the paths are formed by regions that are not connected to each other. This is confirmed by Figure 1(d) that depicts the histogram of this map, exhibiting periodically spaced wavelength ranges of ~100 pm that are not totally reachable with the heaters. We performed additional measurements on a smaller region of the map centered on one of those wavelength ranges in order to better understand the behavior of the tuning mechanism, as depicted on Figure 2. Figure 2(a) represents a wavelength map with a 20-nm OSA range (1565-1585 nm) whereas this range was reduced to 3 nm in the case of Figure 2(b) and (c). It is lower than the FSR of both rings and thus excludes the dominant emitted wavelength of the scan that is located one Vernier channel spacing away from the targeted wavelength. Figure 2(b) shows the raw wavelength map whereas Figure 2(c) only shows points of SMSR > 10 dB. One can see that even if there is no path available for crossing this region with high SMSR at the targeted wavelength, the targeted wavelength is dominant over this small OSA range, thus showing that multimodal operation occurs with modes distant one channel spacing away from each other.

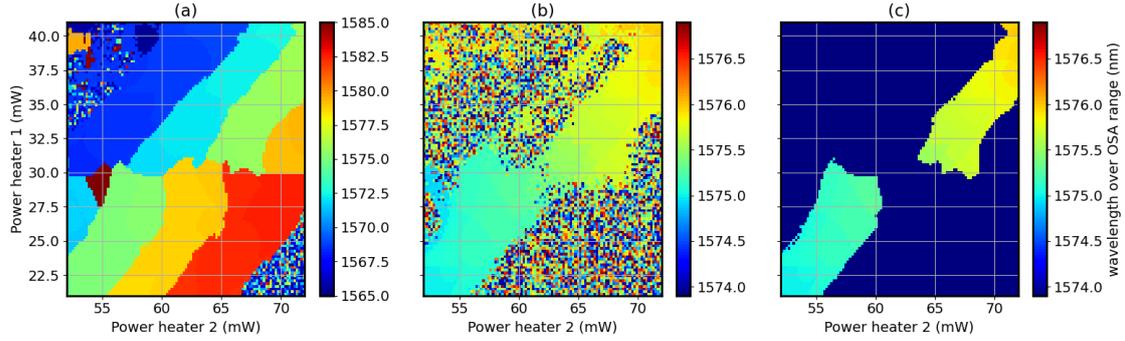

Figure 2. Wavelength maps over a restricted area of 20 mW x 20 mW. (a) Map with OSA range of 20 nm. (b) Map with OSA range of 3 nm. (c) Map (b) with values with SMSR < 10 dB set to zero.

## 3. MODULATION OF EMITTED WAVELENGTH

The phase section is controlled in this paragraph along with Vernier filters in order to investigate for continuous tuning of laser wavelength. The laser diode is connected to an optical isolator as well as a variable optical attenuator (VOA). Signals corresponding to FBG reflection spectra (connection to circulator + FBG) and to interference fringes (connection to Michelson fiber interferometer, MFI) are recorded through an avalanche photodiode connected to an oscilloscope (Keysight DSOX3054T).

### 3.1 Modulation over a small wavelength range and acquisition of FBG spectra

A ramp in dissipated power is applied here on both rings in order to cross a region of interest mentioned above. A similar ramp is applied as well on the phase section and its slope is adjusted in order to evaluate for the tuning ratio ensuring a mode-hop-free tuning of wavelength. After calibration with MFI, reflection spectra of two FBGs were recorded at 50 Hz for a ~400 pm sweep. Figure 3(a) and (b) compare those spectra with spectra obtained from OSA and broadband source. The signals were appropriately stretched so that they superimpose well with OSA spectra.

### 3.2 Modulation over map diagonal at moderate frequency

We then intent to extend the modulated wavelength range by crossing a map diagonal described in 2). Figure 3(c) depicts the signal recorded at a 10-Hz sweep rate with MFI. A phase section ramp is applied for each above mentioned region and its slopes and offsets are adjusted in order to obtain interference fringes over the largest possible range. Here, the interfringe distance is ~210 pm and the number of recorded fringes is in accordance with the wavelength range corresponding to each region of interest.

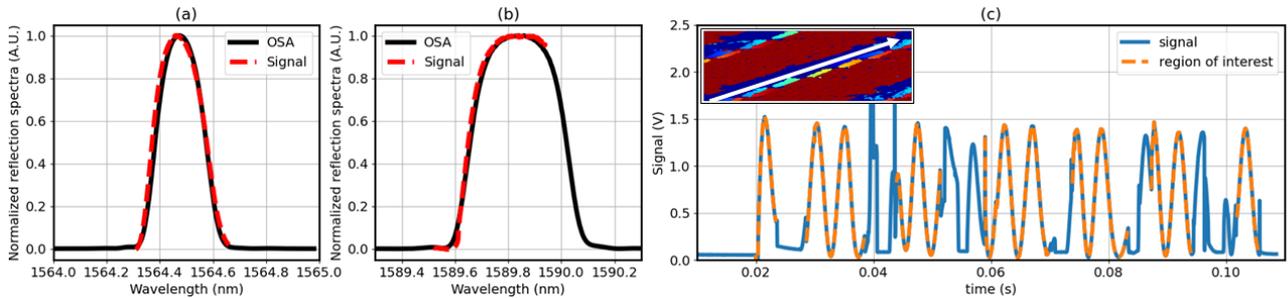

Figure 3. (a) and (b): Comparison of FBG reflection spectra acquired with OSA (black) and single photodiode (red) for two different FBGs. (c) Recorded interference fringes over ~4 nm at 10 Hz. Inset: path on wavelength map.

## 4. DESIGN IMPROVEMENTS FOR A COMPACT INTERROGATION UNIT

Improvements to reach the entire wavelength range of the device might include advanced Vernier designs[6] in order to increase the Vernier gain margin or adjusting ring resonator spectral selectivity through coupling coefficient tuning. However, modulation speed is limited here due to the tuning mechanism based on heaters. Tuning mechanisms based on carrier injection[7,8] are being investigated to speed-up modulation. A compact PIC-based interrogation unit may also include a highly accurate, unambiguous and broadband integrated wavemeter[9] in order to calibrate emitted wavelength during sweeps.

## 5. CONCLUSION

An experimental study of a III-V-on-Si tunable laser diode was exposed in this paper. Wavelength map analysis showed that the studied diode suffers from wavelength bands that are not reachable with heaters and where the diode emits one channel spacing away from the targeted wavelength. Modulation of emitted wavelength was performed in order to produce FBG reflection spectra at 50 Hz over 400 pm and interference fringes over 4 nm at 10 Hz. Design upgrade may ensure continuous tuning over the entire tuning range and advanced tuning mechanism designs based on carrier injection are being investigated in order to reach high scan rates > 10 kHz.